\documentclass{article}

\usepackage{arxiv}
\usepackage[authoryear]{natbib}

\usepackage{fancyhdr,graphicx,amsmath,amssymb}
\usepackage{algorithm, algorithmic}
\usepackage{url}
\usepackage{hyperref}
\include{pythonlisting}
\usepackage{caption}
\usepackage{subcaption}

\usepackage{setspace} 
\usepackage{lineno}
\usepackage[utf8]{inputenc} % allow utf-8 input
\usepackage[T1]{fontenc}    % use 8-bit T1 fonts
\usepackage{hyperref}       % hyperlinks
\usepackage{url}            % simple URL typesetting
\usepackage{booktabs}       % professional-quality tables
\usepackage{amsfonts}       % blackboard math symbols
\usepackage{nicefrac}       % compact symbols for 1/2, etc.
\usepackage{microtype}      % microtypography
\usepackage{lipsum}		% Can be removed after putting your text content
\usepackage{graphicx}
\usepackage{natbib}
\usepackage{doi}

\title{Deep Integrated Pipeline of Segmentation Guided Classification of Breast Cancer from Ultrasound Images}

\author{ \href{https://orcid.org/0000-0003-4765-9542}{Muhammad Sakib Khan Inan}\thanks{Corresponding Author} \\
	Department of Computer Science and Engineering\\
	East Delta University\\
	Chittagong, Bangladesh \\
	\texttt{sakib.khaninan@gmail.com} \\
	\And
	{Fahim Irfan Alam} \\
	Department of Computer Science and Engineering\\
	University of Chittagong \\
	Chittagong, Bangladesh \\
	\texttt{fahim@cu.ac.bd} \\
	\AND
	Rizwan Hasan \\
	Department of Computer Science and Engineering\\
	East Delta University \\
	Chittagong, Bangladesh \\
	\texttt{rizwan.hasan486@gmail.com} \\
}

\renewcommand{\headeright}{Accepted Manuscript}
\renewcommand{\undertitle}{Accepted For Publication in Biomedical Signal Processing and Control, Elsevier}
\renewcommand{\shorttitle}{\textit{Deep Learning Pipeline for Automated Detection of Breast Cancer}}

\begin{document}
\maketitle

\begin{abstract}
Breast cancer has become a symbol of tremendous concern in the modern world, as it is one of the major causes of cancer mortality worldwide. In this regard, breast ultrasonography images are frequently utilized by doctors to diagnose breast cancer at an early stage. However, the complex artifacts and heavily noised breast ultrasonography images make diagnosis a great challenge. Furthermore, the ever-increasing number of patients being screened for breast cancer necessitates the use of automated end-to-end technology for highly accurate diagnosis at a low cost and in a short time.  In this concern, to develop an end-to-end integrated pipeline for breast ultrasonography image classification, we conducted an exhaustive analysis of image preprocessing methods such as K Means++ and SLIC, as well as four transfer learning models such as VGG16, VGG19, DenseNet121, and ResNet50.  With a Dice-coefficient score of 63.4 in the segmentation stage and accuracy and an F1-Score (Benign) of 73.72 percent and 78.92 percent in the classification stage, the combination of SLIC, UNET, and VGG16 outperformed all other integrated combinations. Finally, we have proposed an end to end integrated automated pipelining framework which includes preprocessing with SLIC to capture super-pixel features from the complex artifact of ultrasonography images, complementing semantic segmentation with modified U-Net, leading to breast tumor classification using a transfer learning approach with a pre-trained VGG16 and a densely connected neural network. The proposed automated pipeline can be effectively implemented to assist medical practitioners in making more accurate and timely diagnoses of breast cancer.
\end{abstract}

% keywords can be removed

\section{Introduction}
Cancer, a term that has been continuously threatening humans for decades, is a disease in which cells in one place of the body proliferate and replicate uncontrollably. The cells affected by cancer can penetrate and kill other healthy cells in the human body. It is generally engendered in one part of the human body, and via a process called metastasis, it can spread all over the human body \citep{nhs-website-2021}. Breast Cancer is one of the most frequent types of cancer that has a negative impact on the lives of women of all ages all over the world \citep{nhs-website-2021B}. According to reports, a woman in the United States has a 13\% chance of acquiring breast cancer at some point in her life \citep{ACS}. Breast cancer is estimated to kill 43,600 people in the United States in 2021. Breast Cancer may affect males as well as women. In 2021, around 2,650 new instances of invasive breast cancer in males are predicted to be diagnosed. A man's lifetime chance of developing breast cancer is around 1 in 833 \citep{US-Breast-Cancer-Statistics}.  Breast cancer risk factors include advancing age, obesity, heavy alcohol use, a family history of breast cancer, a history of radiation exposure, a reproductive history, cigarette use, and postmenopausal hormone treatment \citep{WHO}. However, according to the World Health Organization (WHO), half of all breast cancers occur in women who have no recognized risk factors for the disease other than being female and being above the age of 40.
\\

Early detection of breast cancer is very crucial for saving humans in today's world. Doctors typically recommend a breast ultrasonography test to determine whether or not a person has a breast tumor. Generally, breast tumors are of two types: malignant (cancerous cells) and benign (non-cancerous cells). Malignant and benign tumors have distinct morphology in terms of form and texture. As a result, doctors can generally discover the kind or existence of breast cancer by analyzing the tumor's appearance. However, the number of people getting screened for breast cancer is steadily growing. With manual reporting, medical practitioners are facing a great challenge to diagnose this large number of patients. As a consequence, medical practitioners urgently require the automation of the process of detecting breast cancer by reviewing test data such as breast ultrasonography images. \\ 

This scenario has piqued the interest of artificial intelligence (AI) experts, and so AI researchers have committed themselves to investigate the potential capabilities of state-of-the-art machine learning and deep learning approaches for automatically diagnosing breast cancer \citep{mckinney2020international}. In recent, Artificial Intelligence (AI) based methods evinced great potential capability of detecting breast cancer with high accuracy \citep{pisano2020ai}. However, medical image processing and analysis with computerized technologies is great challenge for the researchers from industry and academia \citep{scholl2011challenges}. And, a series of extensive image processing and computer vision methods is often incorporated for successfully processing medical images like breast ultrsonography image, to extract meaningful information from that ultrasonography image \citep{razzak2018deep}. Artificial Intelligence researchers have developed several state of the art pre-processing methods adopting the biomedical imaging domain to tackle challenging artifacts of biomedical images over the years \citep{altaf2019going}. Also, deep learning based image segmentation for medical images have been rigorous investigated by AI researchers to complement the detection of malignant tumor cells \citep{hesamian2019deep}. In particular, breast ultrasonography images are highly noised and posses a complex artifact which requires extensive pre-processing before performing deep learning based segmentation tasks \citep{chen2005classification}. Several researchers have developed and incorporated state of the art image segmentation methods for breast tumor segmentation from raw breast ultrasonography images \citep{saeed2020survey, kim2021artificial}. Though the developed segmentation methods are quite useful for breast cancer detection from breast ultrasonography images, only segmentation of breast tumor area or shape from raw breast ultrasonography images does not directly determine breast cancer malignancy. For this, a number of AI researchers again developed deep learning based individual modules to classify breast cancer from raw breast ultrasound images \citep{qi2019automated}. However, due to the complex artifacts of breast ultrasonography images, using deep learning methods directly to raw ultrasonography images appears to be insufficient. Furthermore, previous studies has addressed the issue by designing modules to improve a particular aspect of the overall breast cancer diagnosis process based on ultrasound images. As a result, these frameworks do not provide a fully integrated pipeline that encompasses preprocessing, segmentation, and classification in a single framework. And so, there is a considerable barrier to the adoption of these frameworks for industrial-level applications \citep{weese2016four}.

A significant research question emerges at this moment, which is \textit{"RQ: Can we fully automate the process of detecting breast cancer from ultrasonography images employing suitable deep learning modules in an integrated way ready for adoption in industry-based medical applications?"}. To answer the aforementioned research question, we effectively used the breast cancer research domain throughout this study and investigated the prospective capability of cutting-edge image processing and deep learning modules when combined in a single end-to-end pipeline. The main contributions of this study are:
\begin{itemize}
    \item We proposed a computer aided deep integrated pipeline of breast ultrasound image segmentation leading to classification for breast cancer diagnosis at an early stage in low cost.
    \item Explored the challenges of processing and analyzing highly noised Breast Ultrasonography images through state of the machine learning and deep learning methods.
    \item Integration of Simple Linear Iterative (SLIC) based unsupervised image segmentation as part of preprocessing to support semantic segmentation by Modified U-Net.
    \item Rigorous analysis of performance and election of best suited pretrained image classification model for integration into our proposed deep integrated model as part of feature extraction in the classification module adopting the domain of medical imaging.
\end{itemize}

The remainder of this paper is structured as follows. Section 2 contains information on the related works. Section 3 introduces the complete methodology including all material and methods which have been utilized in this study. Section 4 contains a rigorous performance analysis of our proposed and experimented integrated deep learning frameworks. Section 5 contains discussion of the research findings.
Finally, section 6 highlights the major conclusions and future work.

\section{Literature Review}\label{literature-review}
In this section, we briefly discuss the previous state of the art studies on breast cancer diagnosis utilizing artificial intelligence based method followed by highlighting the necessity and novelty of our proposed integrated framework in the domain of breast cancer. \\ 

Due to the immense potential of analyzing complex and sensitive data-intensive problems, machine learning and deep learning algorithms have been widely used in several medical applications to tackle challenging medical conditions in recent \citep{shoeibi2020automated}.  \citet{WANG2020105941} proposed a random forest based multi-objective rule extraction method for diagnosing Breast Cancer by utilizing Wisconsin breast cancer datasets. \citet{Inan2021} proposed a hybrid machine learning based method, integrating probabilistic predictions from individual machine learning models with Extreme Gradient Boosting (XgBoost).  \citet{Islam2020} explored the potential of several machine learning algorithms and artificial neural network for breast cancer prediction.
\citet{Ghosh2021} found in their study that RNN (Recurrent Neural Network) based deep learning architectures, the LSTM (Long Short Term Memory) and Gated Recurrent Unit (GRU) outperformed in the diagnosis of breast cancer. Furthermore, several other artificial intelligence based researchers have performed extensive studies for prediction of breast cancer by incorporating state of the art machine learning methods \citep{healthcare8020111, jpm11020061}.  
\\

Medical imaging plays significant role of diagnosis of sensitive medical conditions. Deep learning-based algorithms have recently been incorporated into medical imaging by AI researchers in order to solve sensitive medical situations effectively and frequently \citep{shoeibi2021applications, KHODATARS2021104949}.
Breast Ultrasonography, Histopathology, and MRI are some of the diagnostic tests that doctors use to diagnose breast cancer at an early stage. Breast Ultrasonography is the most commonly recommended test by doctors for the early diagnosis of breast cancer. However, machine learning based methods are not sufficient enough to process complex medical images to diagnose Breast Cancer with high efficiency. Deep Learning based methods can corroborate the study of medical images to detect cancers including Breast Cancer. In this regard, deep learning algorithms have been employed to process biomedical image data to detect Breast Cancer in several studies. \citet{TOGACAR2020123592} developed a Convolutional Neural Network (CNN) model to diagnose Breast Cancer from Histopathology  images of Breast Tissue obtained through fine needle aspiration. A residual net (ResNet) based model of 152 layers have been proposed by \citet{gour2020residual} to classify Breast Histopathology Images. Hybrid combination of Convolutional Neural Networks (CNN), Ridge Regression and Linear Discriminant Analsis have been incorporated by \citet{TOGACAR2020109503} to diagnose Breast Cancer from Histopathological Images of Breast Tissues with integrated autoencoders for pre-processing. A Multiparametric Magnetic Resonance Imaging (mpMRI) based Breast Cancer diagnosing with the incorporation of deep learning methods have been studied by a group of researchers \citep{hu2020deep}. \citet{RASTI2017381} implemented a framework that ensemble convolutional neural network to support computer aided diagnosis of Breast Cancer from breast dynamic contrast-enhanced magnetic resonance imaging (DCE-MRI).  \citet{GAO201853} utilized the combination of state of the are deep CNN (Convolutional Neural Network) as a feature generator and  Shallow CNN for synthesizing contrast-enhanced digital mammography (CEDM) to support the diagnosis of Breast Cancer.However, MRI (Magnetic Image Resonance) based diagnosis are costly and time consuming \citep{SHOEIBI2021104697, sadeghi2021overview}. \\

For this, medical practitioners mostly prescribe Breast Ultrasound Test for an early and rapid diagnosis of Breast Cancer. In this regard, artificial intelligence researcher have extensive working with Breast Ultrasonography images involving both segmentation and classification tasks for early diagnosis of Breast Cancer. Integration of pre-trained state of art Conovlutional Neural Networks  have been proposed by \citet{Masud2021} for the classification of Breast Cancer from Breast Ultrasonography images. 
Breast Ultrasonogrphy images are highly noised and contains a very complex artifact which makes it tougher for the deep learning and machine learning methods to learn the precise morphology of breast tumors to support the diagnosis process \citep{LASSAU2019199, Noble2006, SUDARSHAN201697}. In this regard, noise removal of Breast Ultrasound Images to support an effective detection of Breast Cancer using deep convolutional neural networks have been proposed by \citet{Latif2020}. \citet{Hijab} integrated pretrained image classification models utilizing the concept of transfer learning for an effective classification of Breast Cancer from Breast Ultrasound Images. To improve the performance of Breast Cancer diagnosis from image data researchers also have taken multi modal approach combining Mammography images with Ultrasound Images with a selective ensemble classification methods \citep{Cong2017}. \\

Semantic segmentation is conducted by researchers to extract region of interest from medical images to capture the shape and texture of tumor or infected area for diagnostic purpose \citep{ozturk2020skin, ozturk2019convolutional}. Moreover, segmentation of Breast Ultrasound Images is a very crucial and challenging problems which supports the diagnosis of Breast Cancer in a very effective manner. \citet{Lee2020} have proposed a state of the art channel attention module with multi-scale grid average pooling for Ultrasound Image Segmentation of Breasts. \citet{Almajalid} developed a U-Net based deep learning architecture for the segmentation of of Breast Ultrasonography images. Several other machine learning and deep learning researchers have extensively studied integration of computer aided automated artificial intelligence technologies for Breast Ultrasound Image Segmentation \citep{Huang2018, XU20191,Amiri2020}. \\

However, according to our research no study have proposed a fully automated and compact deep integrated pipeline which includes segmentation leading to classification of Breast Cancer from Breast Ultrasound Images. In this study, we have adopted the Breast Cancer domain at every stage of our research framework and handled the challenging artifacts, high noise, inter class similarities of Breast Ultrasonography Images by developing an efficient fully automated deep integrated pipeline which can be adopted to support the medical practitioners in the diagnostic process of Breast Cancer in a more effective and faster way. The proposed integrated end to end compact deep learning pipeline is highly efficient compact solution for the adoption in the medicals and cancer research domain.

\section{Material and Methods}
\begin{figure}[h!]
	\centering
	\includegraphics[scale=0.25]{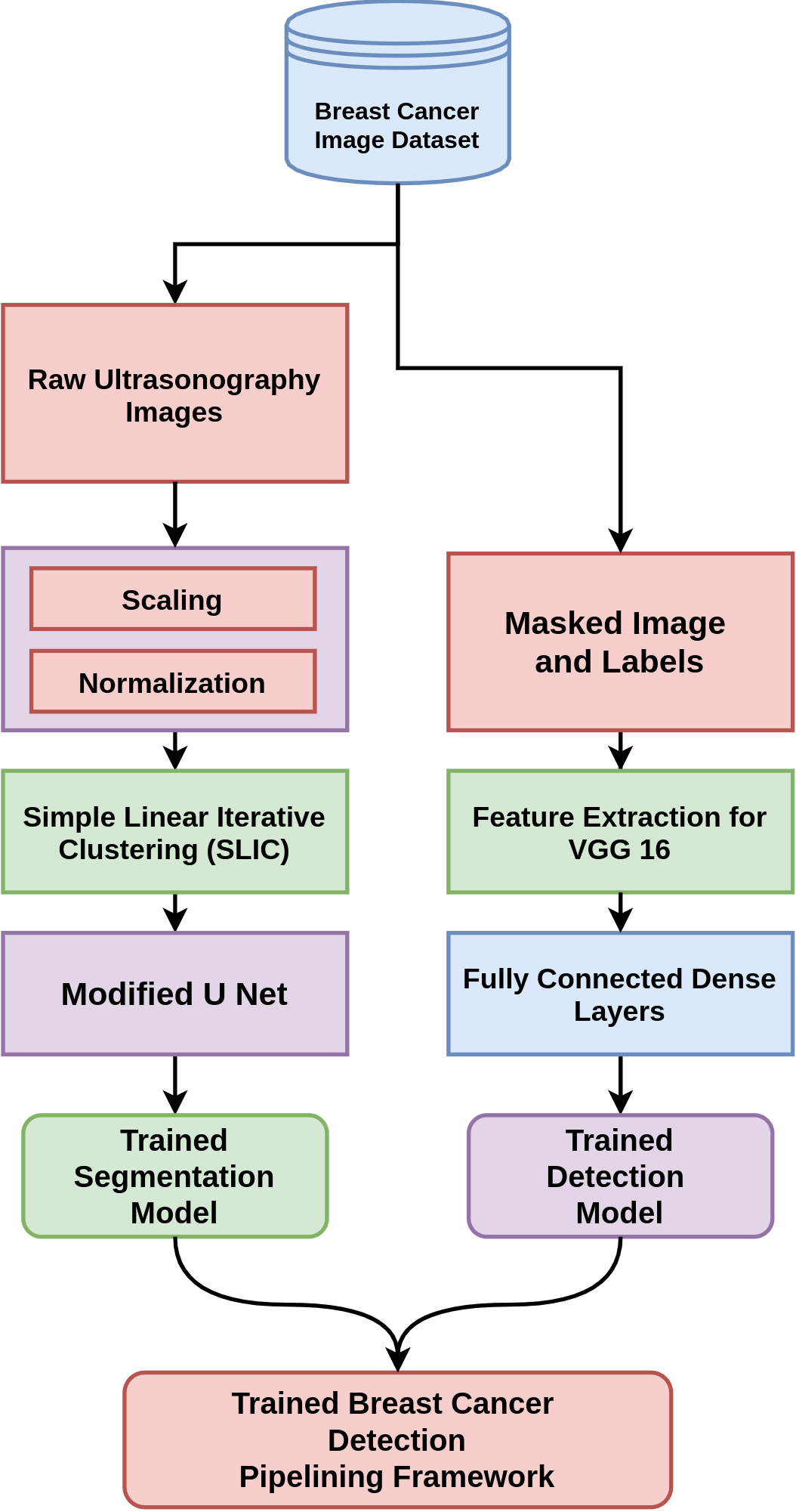}
	\caption{Training Pipeline of Our Proposed Automated Deep Learning Framework}
	\label{fig:training-pipeline}
\end{figure}

In this section, we briefly describe our proposed automated pipelining framework followed by the approaches utilized in our proposed automated deep learning pipeline with illustrations of the individual components that make up the automated framework. At every stage of our research, we have successfully adopted the bio-medical domain with robustness and clarity. A graphical representation of the complete training pipeline of our proposed integrated model is depicted in Figure \ref{fig:training-pipeline}.

\subsection{Proposed Automated Deep Learning Framework}
In this study, we proposed a fully automated end to end pipeline which includes pre-processing with SLIC (Simple Linear Iterative Clustering), semantic segmentation with modified U-Net leading to classification of Breast Cancer with an integrated deep learning framework including pre-trained state-of-the-art feature extraction model.
%%%%%%%% Prediction Pipeline Diagram %%%%%%%%%%%%%%
\begin{figure*}[h!]
	\centering
	\includegraphics[scale=0.15]{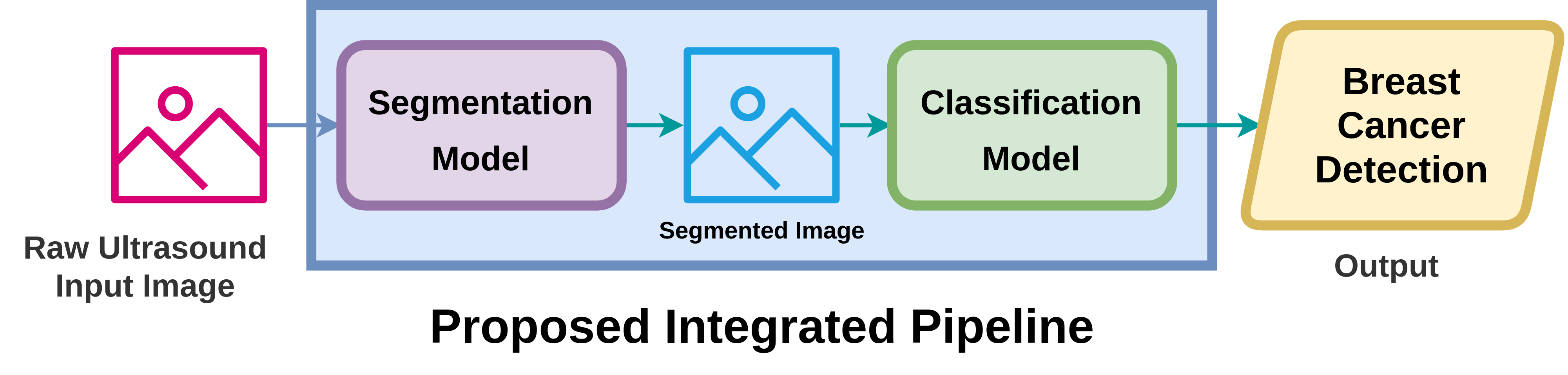}
	\caption{Breast Cancer Diagnosis Framework From Breast Ultrasonic Image}
	\label{fig:prediction-pipeline}
\end{figure*}
%%%%%%%% Prediction Pipeline Diagram %%%%%%%%%%%%
For the training stage, as delineated in the Training Algorithm \ref{algo-train}, the ultrasonography image of breasts have been pre-processed into 128x128x3 pixels. Later, by using unsupervised image segmentation method SLIC (Simple Linear Iterative Clustering), the ultrasonography images are clustered into superpixels that provides a better representation for further semantic segmentation task. All the training images that have been clustered using SLIC is then feed to our modified U-Net architecture for semantic segmentation along with their correspoiding Ground Truth Masks.

\begin{algorithm}[h!]
\caption{Training Algorithm}
	\label{algo-train}
	\hspace*{\algorithmicindent} \textbf{Input:} $\mathbb{R}$ and $\mathbb{M}$ are set of vectors of containing numerical matrix representations of raw ultrasound images and Mask (Ground Truth) images accordingly.
	$\mathbb{Y}$ is set of vector containing image labels (class) for each image in $\mathbb{R}$ and $\mathbb{M}$ consecutively. 
	
	\begin{algorithmic}[1]
	
		\FOR{Each $R_i$ in $\mathbb{R}$} 
		\STATE Apply scale transformation and convert it 128x128 pixels image.
		\STATE Perform unsupervised image segmentation.
		\STATE Normalize the image pixels.
		\ENDFOR
		
		\FOR{Each $M_i$ in $\mathbb{M}$} 
		\STATE Apply scale transformation and convert it 128x128 pixels image.
		\STATE Normalize the image pixels.
		\ENDFOR
		\STATE Define the U-NET architecture for semantic segmentation.
		\STATE Initialize, $Epoch$ to 1,
		\WHILE{$Epoch$ <= 300}
		\STATE Set input data as $\mathbb{R}$ to predict $\mathbb{M}$.
		\STATE Train the model on input data to perform Semantic Segmentation.
		\STATE Compute Loss.
		\STATE Update weights according to learning rate.
		\STATE Update $Epoch$ by 1.
		\ENDWHILE
		
		\STATE Define the classification model architecture. (Pre-trained Convnets for Feature Extraction followed by 5 Fully Connected Layers).
		
		\STATE Initialize, $Epoch$ to 1,
		\WHILE{$Epoch$ <= 100}
		\STATE Set input data as $\mathbb{M}$ to predict $\mathbb{Y}$.
		\STATE Extract Features with Pre-trained models from training input data.
		\STATE Train the Fully Connected Layers by extracted features as input.
		\STATE Compute Binray Cross Entropy Loss
		\STATE Update weights of Fully Connected Layers according to learning rate.
		\STATE Update $Epoch$ by 1.
		\ENDWHILE

	\end{algorithmic}
	\hspace*{\algorithmicindent} \textbf{Result:} Trained Segmentation Model and Classification Model.
\end{algorithm}

After training the modified U-Net for creating semantic segmentation of Breast Ultrasound Images to predict Masked Segmented Image, we trained our integrated classification model which contains pre-trained VGG 16 based feature extraction from Masked Breast Cancer Images to predict whether the Masked Image represent a Malignant Breast Cancer or not. The classification model was trained for 100 Epochs and best weights based lower validation loss was extracted for final modelling.

At the end of training stage, our individual deep learning models for semantic segmentation of Breast Ultrasound Images and classification of Breast Cancer category is ready to utilize. In this context, we designed our proposed automated deep learning pipeline for breast ultrasound image classification which is presented in Algorithm \ref{algo-test}.
\begin{algorithm}[h]
\caption{Automated Deep Learning Pipeline for Breast Ultrasound Image Classifications}
	\label{algo-test}
	\hspace*{\algorithmicindent} \textbf{Input:} $\mathbb{X}$ is set of vectors of containing numerical matrix representations of raw ultrasound images.
	
	\begin{algorithmic}[1]
	\STATE Create a vector $\mathbb{T}$ for storing predicted mask images by segmentation model.
	\FOR{Each $X_i$ in $\mathbb{X}$} 
		\STATE  Apply scale transformation and convert it 128x128 pixels image.
		\STATE Perform unsupervised image segmentation.
		\STATE Normalize the image pixels.
		\STATE Predict the Segmented Mask image of the corresponding $X_i$ by previously trained segmentation model and save it in the vector $\mathbb{T}$.
		\ENDFOR
	\FOR{Each $T_i$ in $\mathbb{T}$} 
		\STATE Apply scale transformation and convert it 128x128x3 pixels image.
		\STATE Predict the classification label of the transformed image by the previously trained classification model.
		\ENDFOR
	\end{algorithmic}
	\hspace*{\algorithmicindent} \textbf{Output:} Classification of Breast Cancer. (Benign/Malignant/Normal)
\end{algorithm}

The integrated deep learning pipeline proposed in this study, would take a 128x128x3 pixels Breast Ultrasonography Image as an input. The input image would be pre-processed using SLIC based unsupervised clustering. Then, using our trained modified U-Net model, a segmented masked image would be generated. The segmented masked followed by necessary preprocessing steps would be incorporated into our trained classification model which will predict the category of that Breast Tumor intro three classes including, Benign, Malignant and Normal. A simple graphical illustration of our fully automated prediction pipeline has been depicted in Figure \ref{fig:prediction-pipeline}. It is clearly seen from the prediction pipeline of Figure \ref{fig:prediction-pipeline} that the pipeline just need the raw breast ultrasonography images, from that it can automatically generate masked images leading to final classification of Breast Cancer.

\subsection{Medical Image Segmentation}
Segmentation in medical imaging plays a critical role in the detection, localization, or identification of shapes of tumor cells. Cancer severity is determined by the morphology of the malignant cells, which may be determined using image classification algorithms using the shape of cancerous cells detected from image segmentation. The necessity of ultrasound image segmentation has been investigated by several researchers earlier \citep{Noble2006, Chauhan2010, SHEELA2016}. In general, there are two types of image segmentation in a broader sense and is Unsupervised Segmentation which refers to segmentation without using labeled image pixels, and Semantic or Supervised Segmentation which refers to segmentation using labeled training images. In our study, we incorporated a hybrid integration of the Unsupervised and the Supervised image segmentation methods to successfully classify Breast Cancer from Breast Ultrasound Images.

\subsubsection{Unsupervised Segmentation}
In recent, unsupervised segmentation has been utilized for medical image segmentation in several state of art studies with great efficiency \citep{Aganj2018}. It is useful when the training data is not labeled for the segmentation task \citep{Baby2020}.  However, processing ultrasound images are very challenging due to their complex image artifacts \citep{LIU2019}. In our study, we have experimented with two widely used unsupervised medical image segmentation methods (K Means++ and SLIC) as a part of image pre-processing to ameliorate the efficiency of the semantic segmentation to generate segmented images of breast tumors.

\textbf{\textit{K-Means++}}

\begin{figure*}[h!]
\centering
\begin{subfigure}{0.4\textwidth}
  \centering
  % include first image
  \includegraphics[scale=0.4]{figs/Org_Benign_12.png}  
  \caption{Original Image (Benign Tumor)}
  \label{fig:k-org}
\end{subfigure}
\begin{subfigure}{0.4\textwidth}
  \centering
  % include second image
  \includegraphics[scale=0.4]{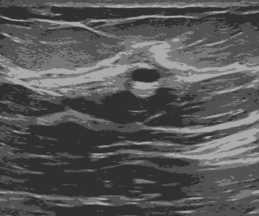}  
  \caption{K-Means++ Segmented Image (Benign Tumor)}
  \label{fig:k-seg}
\end{subfigure}
\caption{Image Segmentation with K-Means++}
\label{fig:kmeans}
\end{figure*}

K Means++ is an unsupervised image segmentation algorithm that is an improved version of the popular K Means clustering algorithm \citep{kmeansppAlgo}. In medical image processing and analysis, K Means clustering has been utilized to segment interest areas from the background for spotting out the location and shape of cancerous cells \citep{kmeansppReview}. In K Means++, the interest area representing a better view to locate and analyze cancerous cells is segmented by inducing clusters or partitions where similar pixels of images are grouped into a certain category through an iterative process. Here, $K$ indicates the number of clusters to consider. 
% The first centroid in the K-Means++ algorithm is chosen randomly and the following centroid is chosen based on probability, which is determined by the distance between the first and second points. The data points are assigned to the group of the closest centroid. The process is iteratively applied until convergence. 
This algorithm is relatively fast, computationally less expensive, and can easily be adapted to new unseen data. The effect of incorporating the K Means++ algorithm for Breast Ultrasound Image Segmentation is illustrated in the Figure ~\ref{fig:kmeans}. Here, Figure \ref{fig:k-org} represents original ultrasound image of Breast for patient having Benign Tumor and Figure \ref{fig:k-seg} shows segmented image of that particular ultrasound image after incorporating K Means++ algorithm.

In this study, we have utilized the OpenCV implementation in Python for K Means++ with a $K$ -value of 4 and stopping criteria of if the maximum iterations are reached or specified epsilon/accuracy is reached. From the Figure, we can notice that the original ultrasound image has been segmented into clusters where the image pixels are stretched to their nearest most significant pixel values with normalization of 0 to 1.

\textbf{\textit{SLIC (Simple Linear Iterative Clustering)}} 

SLIC(Simple Linear Iterative Clustering) is a widely used clustering algorithm used for unsupervised image segmentation that clusters image pixels into nearly uniform superpixels \citep{SLIC-Paper}. A superpixel is a color-based segmentation that can be delineated as a set of pixels sharing similar characteristics. Many studies have used SLIC for image segmentation due to its simplicity and computational efficiency \citep{Xie2018, Chi2013}. SLIC has been proved an efficient method for image segmentation to get meaningful insights from medical images including ultrasonography images \citep{Fang2020, Wang2020}. In our study, for segmentation of Breast Tumor from Breast Ultrasound Images, we have integrated SLIC based image segmentation as part of pre-processing to support the performance of semantic segmentation.  The integration of SLIC may improve the non-linearity issue of ultrasound image pixel distribution by grouping pixels of similar characteristics.

\begin{figure*}[h!]
\centering
\begin{subfigure}{0.4\textwidth}
  \centering
  % include first image
  \includegraphics[scale=0.4]{figs/Org_Benign_12.png}  
  \caption{Original Image (Benign Tumor)}
  \label{fig:SLIC-org}
\end{subfigure}
\begin{subfigure}{0.4\textwidth}
  \centering
  % include second image
  \includegraphics[scale=0.4]{figs/SLIC_Benign_12.png}  
  \caption{SLIC based Segmented Image (Benign Tumor)}
  \label{fig:SLIC-seg}
\end{subfigure}
\caption{Image Segmentation with SLIC}
\label{fig:SLIC}
\end{figure*}

Here, the effect of SLIC based image segmentation is depicted in the Figure ~\ref{fig:SLIC}. In Figure ~\ref{fig:SLIC-seg}, after applying SLIC, the image pixels sharing similarity in distribution are assigned into a particular cluster making the image easy to process and analyze. In this study, for implementation purposes, the OpenCV library has been used with a configuration of region size 20 of super-pixels, smoothing factor of 10 and 100 iterations per image.

\subsubsection{Semantic Segmentation: U-Net}
U-NET is a state-of-the-art deep architecture that was proposed especially in consideration of  Biomedical Image segmentation \citep{Ronneberger2015}. It performs semantic or supervised segmentation.  The U-Net model comprises an Encoder-Decoder architecture. The encoder is basically a contraction path that learns the context in the medical image utilizing stacked convolutional layers and pooling layers (max)  followed by a decoder path with the functionality of symmetric expansion for the localization of tumor cells precisely utilizing transposed convolutions. The network contains only fully connected convolutional layers. Considering the accuracy and simplicity of this architecture, several researchers have incorporated variants of U-Net for a successful segmentation and detection of biomedical images for the study of Breast Cancer \citep{SOULAMI2021, Honghan2021, Negi2020}. In our study, we have incorporated a variant of UNET architecture for semantic segmentation of Breast Tumors Cells from Breast Ultrasonography images. Every block in the encoder section has two 3x3 Conv layers followed by 2x2 pooling(max) layers. At the end of encoder or contraction layers there lies two 3x3 Conv layers and 2x2  Transposed Conv layer leading to decoder layers. The input image for UNET in this study is 128x128x3 (Width x Height x Channel) pixels. And the final segmented image is 128x128x1 pixels in dimensions.
The Keras implementation of Tensorflow is utilized for experimental purposes. In a standard U-Net, the number of convolutional blocks in the contraction and expansion path are equal. In our optimized UNET architecture, we have used half the number of filters in every second convolutional layer than the first Conv layer each convolutional block of contraction path with a motivation of extracting the robust features by restricting the dimensions. Binary Cross Entropy function is incorporated for calculating loss in the training stage of the model and in the validation stage Dice-Coefficient score has also been computed for better justification of performance \citep{Jadon2020}. Mathematically, Binary Cross-Entropy Loss is can be computed as \citep{ruby2020binary};

\begin{equation}\label{binary-loss}
Log loss = - \frac{1}{N} \sum_{i=1}^{N}y_ilog(p(y_i)) + (1-y_i)log(p(y_i))
\end{equation}

Here, $y_i$ denotes the ith actual pixel of Breast Ultrasonography Image and $p(y_i)$ predicted pixel value for that particular instance. And Dice-Coefficient Score can be computed as \citep{moltz2020learning};

\begin{equation}\label{dice-loss}
diceLoss = \frac{2|A\cap B|}{|A|+ |B|}
\end{equation}

Dice Coefficient a popularly used evaluation metric for image segmentation based tasks that measures overlap between two samples. $|A \cap B|$  represents the common elements between sets $A$ and $B$ \citep{jordan-2020}. Eq. \ref{binary-loss} is incorporated for the computation of loss at every epoch for our modified U-Net model. And, Eq. \ref{dice-loss} is utilized to evaluate the performance of U-Net model at every epoch.  \\

\begin{figure*}[h!]
    \centering
    \includegraphics[scale=0.32]{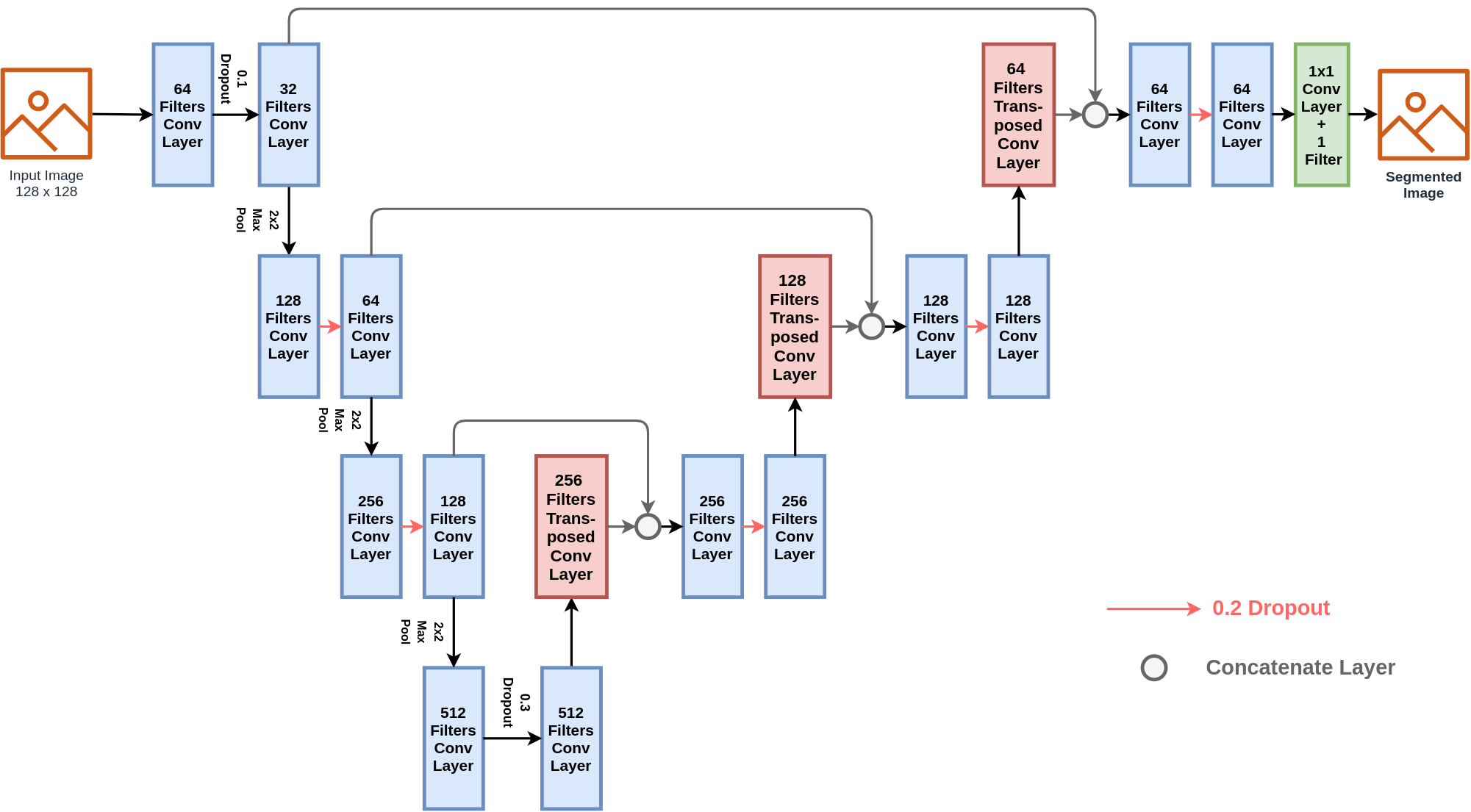}
    \caption{Block diagram of modified U-Net model utilized in our proposed integrated pipeline}
    \label{fig:unet-diagram}
\end{figure*}

A block diagram included with hyper parameters of modified U-Net model is illustrated in Figure \ref{fig:unet-diagram}. In our proposed integrated pipeline, for modified U-Net model, each of the Conv Layers (Convolutional Layer)  comprises of a kernel size 3x3, activation function ReLU (Rectified Linear Unit), 'he\_uniform" as kernel initializer and 'same' padding. However, the Transposed Convolutional Layers comprises of 2x2 kernel size and 2x2 strides each. The Adam optimizer function with a learning rate of 0.0001 is incorporated for optimizing the models performance after each epoch. The popular Tensorflow library for Python programming language has been incorporated for implementation of the aforementioned U-Net segmentation model.
\subsection{Tumor Classification}
In this section, we have briefly discussed deep learning models which have been included in the classification stage of breast cancer detection for experimentation purpose. To corroborate breast cancer classification, we adopted pre-trained models utilizing the potential of transfer learning followed by densely connected fully connected layers.
\subsubsection{Transfer Learning: Feature Extraction with Pre-trained ConvNets}
Pretrained ConvNets (CNNs)  are widely utilized state of the art deep learning models for tasks like Image Classification which have been already trained on the prestigious benchmark dataset ImageNet \citep{deng2009imagenet}. Pretrained ConvNets have immense potential of extracting robust features from biomedical image data without being the need of training it on a unseen dataset \citep{kieffer2017convolutional}. In this study, for classifying Breast Tumor from Breast Ultrasonography Images we have experimented with state of the art pretrained ConvNet(CNN) for robust feature extraction by adopting the Biomedical domain.

\textbf{VGG 16 and VGG 19 :} VGG 16 and VGG 19 \citep{simonyan2014very}, are state of art deep ConvNet architecture which is trained on ImageNet Dataset \citep{deng2009imagenet} for image classification problem. The architecture of VGG 16 and VGG 19 is quite simple and lightweight because of having only small 3x3 conv filters. The 3x3 filters have been designed with the configuration of stride 1 and 'same' padding. A maxpooling layer of 2x2 filters and stride 2 is incorporated in each block after set of convolutional blocks. In our study, we have utilized the pre-trained version of VGG 16 and VGG 19 for extracting features from Breast Ultrasonography Segmented Image at the classification stage. VGG 19 is nearly identical to VGG 16, except it has additional layers and a deeper architecture. VGG 19 contains additional parameters and Convolutational Layers than VGG 16. The VGG  models is useful for object localization because to its tiny filter size-based design. This makes it a strong option for Breast Tumor localisation from Breast Ultrasonography pictures, which is critical for assessing Breast Tumor shape. In our study, the Convolutional Blocks of VGG 16 has been utilized to extract feature by pretrained weights.

\textbf{DenseNet121 :} DenseNet (Dense Convolutional Network) \citep{huang2017densely} is an architecture that focuses on deepening deep learning networks while also making them more cost-effective to train by using shorter connections between layers. The DenseNet architecture is designed to allow maximum information sharing between different network levels. Each layer receives "collective knowledge" from the levels above it. It requires fewer parameters than typical convolutional neural networks since it does not need to train irrelevant feature mappings. DenseNet features two crucial components in addition to the core convolutional and pooling layers. They're referred to as Dense Blocks and Transition Layers. In this study, we experimented with pretrained DenseNet121, which includes 6,12,24,16 layers in the four dense blocks model, to extract robust features for Breast Cancer Detection from Breast Ultrasonography Images, taking into account the vast potential of DenseNet based models. 

\textbf{ResNet50 :} ResNet-50  \citep{He_2016_CVPR} is a convolutional neural network that is 50 layers deep. ResNet50 is a variant of ResNet (Residual Net) \citep{He_2016_CVPR} model which has 48 Convolution layers along with 1 MaxPool and 1 Average Pool layer. It has 3.8 Billion Floating points operations. It is a widely used ResNet model and we have explored ResNet50 architecture in depth. In deep learning, deeper model are generally hard to optimize which effects the performance of deep models with more layers. ResNet utilizes network layers to learn residual mappings instead of directly trying to learn a underlying mapping. The ResNet model solves the previous optimization problems with deeper neural networks. In this study, we have incorporated the ResNet50 model as a pretrained base for feature extraction to diagnose Breast Cancer from Breast Ultrasound Images for experimentation purpose.

\subsubsection{Fully Connected Dense Layers: Classification of Breast Tumor}

Fully Connected Layers, also known as Dense Layers, are a sort of Neural Network in which every node in the previous layer is connected to every node in the following layer, and the outputs of previous layers are used as inputs for the next layers. 
\begin{figure*} [h!]
    \centering
    \includegraphics[scale=0.6]{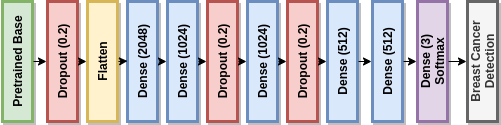}
    \caption{Classification Model of Breast Cancer Detection}
    \label{fig:fc}
\end{figure*}
In this study as depicted in Figure \ref{fig:fc}, we used 5 fully connected layers for classification, with a Dropout of 0.2 between two dense layers of 1024 Neurons to minimize overfitting. For each dense layer, the widely known and efficient 'tanh' activation function was utilized. The 'tanh' activation function is generally computed as Eq. \ref{tanh} ~ \citep{kalman1992tanh} ,
\begin{equation}\label{tanh}
    f(x) = \frac{e^{x} - e^{-x}}{e^{x} + e^{-x}}
\end{equation}

Here, $f(x)$ denotes the activation function where $x$ denotes the input of the function.
At the end of the fully connected layer, a softmax layer with three nodes was utilized to map the class label for detection of breast cancer from breast ultrasonography images.

\section{Experimental Results}
This section contains a thorough depiction of results for various experimental stages in order to effectively illustrate the study's findings. Broadly, our study has three stages of assessment, in terms of pre-processing, segmentation, and classification approaches.

\subsection{Dataset Description}
In this study, to validate the efficiency of our proposed deep integrated pipeline, we have experimented our model on a benchmark dataset of Breast Ultrasonography Images \citep{dataset}. It is a challenging dataset, with women ranging in age from 25 to 75 years old. The collection contains 780 photos, each of which is 500 x 500 pixels in size. PNG files are used for the images. This dataset contains raw and masked ultrasound images of Breast Cancer labeled as Benign (487 samples), Malignant (210 samples) and Normal (177 samples). We have adopted the sensitivity of Biomedical Domain at every stage of our framework and solved the challenges of noisy and complex artifact Breast Ultrasound Images to successfully detect Breast Cancer. For experimentation purpose to ensure the clarity of the results, we have spitted our dataset with a hold-out validation ratio of 80:20 (training:testing) followed by a randomization seed of value 15 through incorporation of popular Scikit-Learn library in Python.
\subsection{Evaluation Metrics}
The choice of evaluation metrics are very important in domain of Artificial Intelligence for Biomedical or Health Care. Considering the sensitivity of Breast Cancer research, we have evaluated our Breast Cancer Detection models efficiency by incorporating below mentioned metrics \citep{furnkranz2003analysis} summarized in Table \ref{tab:metrics-summary}:

\begin{table*}[h!]
\centering
\begin{tabular}{|l|l|}
\hline
\textbf{Metric} &
  \textbf{Description} \\ \hline
True Positive (TP) &
  \begin{tabular}[c]{@{}l@{}}The case when patient is actually suffering from the cancer and the \\ model also classified as positive.\end{tabular} \\ \hline
False Positive (FP) &
  \begin{tabular}[c]{@{}l@{}}The case when patient is not suffering from the cancer but \\ the model classified as positive.\end{tabular} \\ \hline
True Negative (TN) &
  \begin{tabular}[c]{@{}l@{}}The case when patient is not suffering from the cancer and \\ the model also classified as negative.\end{tabular} \\ \hline
False Negative (FN) &
  \begin{tabular}[c]{@{}l@{}}The case when patient is actually suffering from the cancer but \\ the model classified as negative.\end{tabular} \\ \hline
Accuracy &
  \begin{tabular}[c]{@{}l@{}}It defines correctly identified category of Breast Cancer. It is defined as:\\ Accuracy = (TP + TN) / (TP + FP + TN + FN)\end{tabular} \\ \hline
Recall/True-Positive Rate: &
  \begin{tabular}[c]{@{}l@{}}It is defined as the ratio of the number of samples that have been correctly \\ predicted corresponding to all of the samples in the data.  It can define as-\\ Recall = TP / (TP + FN)\end{tabular} \\ \hline
Precision &
  \begin{tabular}[c]{@{}l@{}}It is defined as the ratio of the number of samples that have been correctly \\ predicted corresponding to all samples of the particular category. It can define as-\\ Precision = TP / (TP + FP)\end{tabular} \\ \hline
F1-Score &
  \begin{tabular}[c]{@{}l@{}}It is defined as the term that balances between recall and precision. It can define as-\\ F1-Score = 2 * (Recall * Precision) \textbackslash (Recall + Precision)\end{tabular} \\ \hline
\end{tabular}
\caption{Description of Performance Evaluation Metrics}
\label{tab:metrics-summary}
\end{table*}

\subsection{Evaluation Stage 1: Segmentation}
In this stage of evaluation, we evaluated our segmentation module which is a very important part of our deep integrated pipeline. Segmentation of Breast Ultrasound Images is a very challenging task in the domiain of Computer Vision for Medical Imaging as Ultrasonography Images have complex artifacts, highly noised and posses multi-co-linearity issues \citep{LASSAU2019199}. \\

To solve this issue, in our study we have experimented with two state of art unsupervised image segmentation method K-Means++ and SLIC (Simple Linear Iterative Clustering) as a part of preprocessing Ultrasonography Images to corroborate the performance of Modified-UNET for semantic segmentation. The Modified U-Net model for segmentation was trained for 300 Epochs and the best weights according minimum binary cross entropy loss selected as segmentation model. Table ~\ref{tab:segmentation-results} illustrate the performance of U-Net segmentation in terms of Binary Cross Entropy Loss and Dice-Coefficient Scores on testing dataset. \\

%%%%%%%% U-NET Segmentation Scores %%%%%%%%%%%%%%
\begin{table}[htbp]
\centering
\begin{tabular}{|c|c|c|}
\hline
\textbf{Method} &
  \textbf{\begin{tabular}[c]{@{}c@{}}Binary Cross \\ Entropy Loss\end{tabular}} &
  \textbf{\begin{tabular}[c]{@{}c@{}}Dice Coefficient \\ Score\end{tabular}} \\ \hline
\begin{tabular}[c]{@{}c@{}}SLIC  \& \\ UNET\end{tabular}    & 0.2149 & 63.4  \\ \hline
\begin{tabular}[c]{@{}c@{}}KMeans++ \\ \& UNET\end{tabular} & 0.2547 & 57.31 \\ \hline
UNET                                                        & 0.2595 & 60.67 \\ \hline
\end{tabular}
\caption{Performance Evaluation Scores of Ultrasound Image Segmentation}
\label{tab:segmentation-results}
\end{table}

The high dice-coefficient scores and low binary cross entropy loss score of SLIC (Simple Linear Iterative Clustering) integrated Modified-UNET in comparison to UNET without preprocessing and UNET with K-Means++ validates the relevance of including SLIC based unsupervised segmentation as a part of preprocessing to corroborate Modified-UNET for better semantic segmentation of Ultrasound Images. However, the results from Table \ref{tab:segmentation-results} also illustrates that pre-processing with K-Means++ is not effictive method for Breast Ultrasonography Images as it is worsening the result. From a hypothetical analysis of K-Means++ based unsupervised segmentation of Breast Ultrasonography images, as breast ultrasonography images is highly noised and highly correlated dominant black and white pixel have overlapping issues, it is noticed that K-Means++ is actually clustering the image pixels in a way which is increasing the multi-co-linearity issue in the pixel distribution of Breast Ultrasonography Images. This scenario is also reflected through Figure \ref{fig:kmeans}. This proves K-Means++ inefficient for incorporating it in our proposed architecture as part of pre-processing for Breast Ultrasonography Image Segmentation. It is new finding in the domain of Breast Ultrasonography Image analysis which have been presented in this study, with strong evidence. \\

 \begin{figure*}[h!]
\centering
\begin{subfigure}{\textwidth}
  \centering
  % include first image
  \includegraphics[scale=0.5]{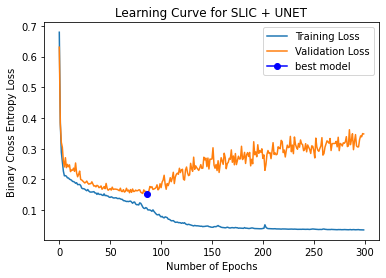}  
  \caption{Learning Curve for U-Net with SLIC in Preprocessing Based on Binary Crossentropy Loss}
  \label{figs:slic-binary-loss}
\end{subfigure}
\newline
\begin{subfigure}{\textwidth}
  \centering
  % include second image
  \includegraphics[scale=0.5]{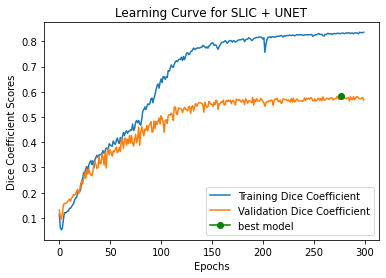}  
  \caption{Learning Curve for U-Net with SLIC in Preprocessing Based on Dice Coefficients}
  \label{fig:slic-dice-loss}
\end{subfigure}
\caption{Semantic Segmentation with UNET with SLIC Preprocessing}
\label{fig:slic-unet}
\end{figure*}

The learning curves of SLIC integrated modified UNET model is illustrated in Figure \ref{fig:slic-unet}. It delineates that the model's performance is increasing with the increasing number of epochs in terms of Dice-coefficient scores for both training images and validation images. The volatility and discrepancy between the learning curves of training and validation data, on the other hand, indicates a slight overfitting problem. This slight overfitting is caused by the fact that we have less training and validation samples because our dataset is smaller in terms of deep learning. Because of their intricacy, deep learning models in general tend to overfit. A slight overfitting can be ignored if the model's overall potential is significant. To ensure minimum overfitting, the best weights based on maximum validation dice-coefficient scores have been finalized for our proposed model at the end of training stage for further experimentation purpose. According to the dice-coefficient scores and binary cross entropy scores, we elected the configuration of pre-processing with Simple Linear Iterative Clustering (SLIC) leading to semantic segmentation with modified U-Net for the integration into our proposed deep integrated pipeline for detection of Breast Cancer from Breast Ultrasonography Images.

\subsection{Evaluation Stage 2: Detection of Breast Cancer}
In this stage of evaluation, we have investigated the performance of our proposed deep integrated model in comparison to other possible state of the art configurations.  Table \ref{tab:slic-unet-all} delineates the performance of state of the art transfer learning models for in terms popular evaluation metrics including Precision, Recall and F1-Scores for three classes Benign, Malignant and Normal.

\begin{table*}[h!]
\centering
\begin{tabular}{|c|c|c|c|c|}
\hline
\textbf{Classifiers} &
  \textbf{\begin{tabular}[c]{@{}c@{}}SLIC+ \\ UNET + VGG 19\end{tabular}} &
  \textbf{\textbf{\begin{tabular}[c]{@{}c@{}}SLIC + \\ UNET + VGG 16\end{tabular}}} &
  \textbf{\begin{tabular}[c]{@{}c@{}}SLIC + \\ UNET + ResNet50\end{tabular}} &
  \textbf{\begin{tabular}[c]{@{}c@{}}SLIC + \\ UNET + DenseNet121\end{tabular}} \\ \hline
% Loss          & 3.27    & 2.74    & 1.13    & 1.8     \\ \hline
% Accuracy      & 69.87\% & 73.72\% & 69.23\% & 71.15\% \\ \hline
% AUC Score     & 80\%    & 83\%    & 85\%    & 84\%    \\ \hline
Precision (B) & 74.74\% & 74.49\% & 69.73\% & 74.74\% \\ \hline
Recall (B)    & 81.61\% & 83.91\% & 87.36\% & 81.61\% \\ \hline
F1-Score(B)   & 78.02\% & 78.92\% & 77.51\% & 78.02\% \\ \hline
Precision (M) & 60.00\% & 72.50\% & 84.61\% & 62.75\% \\ \hline
Recall (M)    & 61.36\% & 65.91\% & 25.00\% & 72.73\% \\ \hline
F1-Score(M)   & 60.67\% & 69.05\% & 38.60\% & 67.37\% \\ \hline
Precision (N) & 68.75\% & 72.22\% & 61.77\% & 80.00\% \\ \hline
Recall (N)    & 44.00\% & 52.00\% & 84.00\% & 32.00\% \\ \hline
F1-Score(N)   & 53.66\% & 60.47\% & 71.19\% & 45.70\% \\ \hline
\end{tabular}
\caption{Classification Scores of SLIC + UNET + Transfer Learning}
\label{tab:slic-unet-all}
\end{table*}

\begin{figure*}[h!]
    \centering
    \includegraphics[scale=0.5]{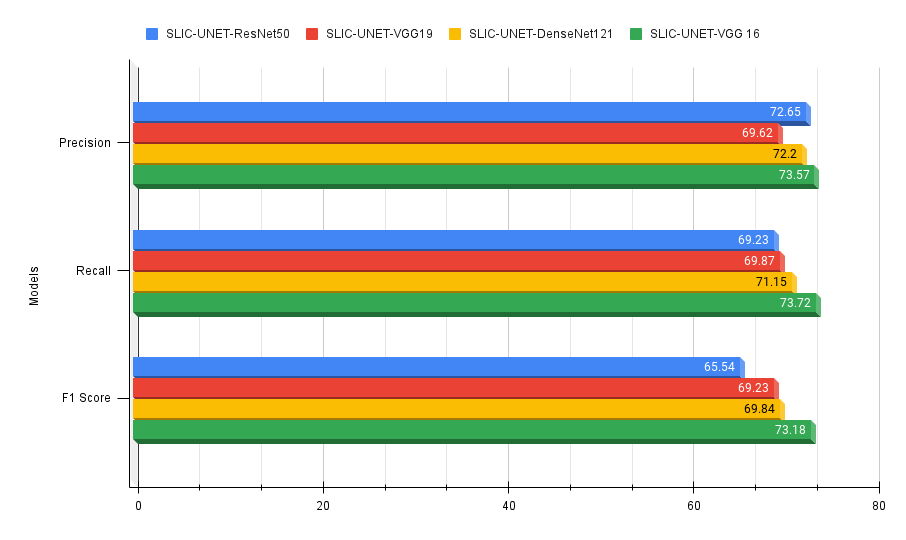}
    \caption{Weighted Average Precision, Recall and F1-Scores of Deep Learning Models for Breast Cancer Detection}
    \label{fig:wavg-scores}
\end{figure*}

By analyzing the scores of evaluation metrics presented in Table \ref{tab:slic-unet-all}, it is evident that the pretrained VGG 16 model with the integration of SLIC and Modified UNET is comparatively outperforming in most of evaluation criteria against other pretrained model used for feature extraction. However, as the dataset is highly imbalanced in terms of class distribution, an efficient model should be able to detect the three classes efficiently with a balanced performance. Considering this, we investigate the Weighted Average of Precision, Recall and F1-Scores to identify the most consistent pretrained model for classification in this stage. The weighted average scores of Precision, Recall and F1-Score is illustrated in the Figure \ref{fig:wavg-scores}.

From Figure \ref{fig:wavg-scores}, it is clearly visible that VGG 16 with the above mentioned settings is out-performing all other state of the art pretrained models with a consistent performance in Weighted Average Scores of Precision, Recall and F1-Score. However, the ResNet50 pretrained model is evincing a competitive score in terms of Weighted Average Precision. But in Weighted Average Recall and F1-Scores the performance of ResNet model is exacerbating. The other two pretrained models DenseNet121 and VGG 19 is also showing consisting performance but not outperforming the performance of VGG 16 model.

\begin{figure*}[htbp]
    \centering
    \includegraphics[scale=0.5]{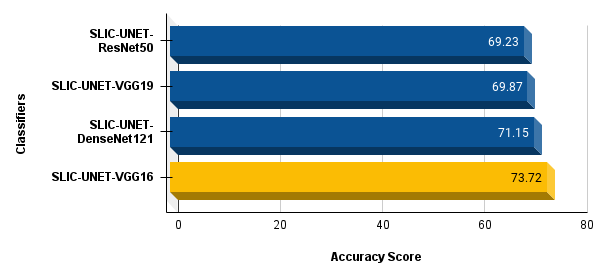}
    \caption{Accuracy Scores of Deep Learning Models for Breast Cancer Detection}
    \label{fig:accuracy-scores}
\end{figure*}

The depiction of Accuracy Score of integrated combinations have been depicted in Figure \ref{fig:accuracy-scores}.

\begin{table*}[H]
\centering
\begin{tabular}{|l|c|c|c|}
\hline
\textbf{Classifiers} & \textbf{SD of F1 Score} & \textbf{SD of Precision} & \textbf{SD of Recall} \\ \hline
% UNET + VGG 19                  & 0.1088 & 0.1906 & 0.0145 \\ \hline
% UNET + VGG 16                  & 0.1060 & 0.2326 & 0.0746 \\ \hline
% UNET + ResNet50                & 0.2353 & 0.1083 & 0.3365 \\ \hline
% UNET + DenseNet121             & 0.0953 & 0.2385 & 0.1097 \\ \hline
SLIC+ UNET + VGG 19            & 0.1254 & 0.0741 & 0.1882 \\ \hline
SLIC + UNET + VGG 16           & 0.0923 & 0.0124 & 0.1600 \\ \hline
SLIC + UNET + ResNet50         & 0.2088 & 0.1160 & 0.3507 \\ \hline
SLIC + UNET + DenseNet121      & 0.1647 & 0.0884 & 0.2645 \\ \hline
% KMeans PP + UNET + VGG 19      & 0.0766 & 0.1305 & 0.0212 \\ \hline
% KMeans PP + UNET + ResNet50    & 0.2828 & 0.1509 & 0.4106 \\ \hline
% KMeans PP + UNET + DenseNet121 & 0.0712 & 0.1169 & 0.0949 \\ \hline
% KMeans  PP + UNET + VGG 16     & 0.0878 & 0.1640 & 0.0898 \\ \hline
\end{tabular}
\caption{Standard Deviation of F1 Scores, Precision and Recall among Three Classes}
\label{tab:stdev-all}
\end{table*}

To validate the consistency in performance across all three classes (Benign, Malignant and Normal) of Breast Cancer, we computed Standard Deviation of Precision, Recall and F1-Scores for Benign, Malignant and Normal samples, presented in Table \ref{tab:stdev-all}. The low standard deviation scores of VGG 16 model among all classes in comparison to other models, proves it as the most consistent model for Breast Cancer Detection from Ultrasonography Images.

\section{Discussion}
In segmentation stage of the study, we can see from the Table \ref{tab:segmentation-results} that the SLIC integrated U-Net's Dice-coefficient score is 63.4 which about 3 units higher than U-Net without any pre-processing. And so, it is clearly visible from the results presented in Table \ref{tab:segmentation-results} that the integration of Simple Linear Iterative Clustering (SLIC) as a pre-processing step complements the semantic segmentation task with modified U-Net. Moreover, the binary cross entropy loss of SLIC integrated UNET is minimum of all other combinations. This supports the hypothesis that super-pixel-based features can be used to improve the performance of semantic segmentation or pixel-wise classification of breast ultrasound images. As a result, Simple Linear Iterative Clustering (SLIC) has been considered as a strong candidate for incorporation in our proposed deep pipeline. However, only segmentation or pixel-wise classification is not enough automating the whole process of breast cancer diagnosis. And so, we investigated further the potential possibility of integrating these modules into one single framework leading to final diagnosis.  In the final breast cancer detection stage, after rigorous analysis of scores based on standard evaluation metrics, the VGG 16 model along with SLIC and UNET, evinces the capability of most potential combination for integrating it in the domain of Breast Cancer Detection from Breast Ultrasonography Images with a Accuracy of 73.72 \% obtained from Figure \ref{fig:accuracy-scores}, which is significantly higher than other experimented combinations. In biomedical domain, in terms of disease diagnosis, the Recall value is a significant measure which should higher for an efficient artificial intelligence model. From Figure \ref{fig:wavg-scores}, it is evident that the Weighted Average Recall value of SLIC-UNET-VGG16 combination is 73.72\% which is significantly higher than other combinations. Further more, Weighted Average Precision and F1-Score of this combination is slightly higher with a value of 73.57\% and 73.18\% accordingly. It shows that SLIC-UNET-VGG16 can diagnose breast cancer malignancy with greater precision than other combinations, which is a critical criterion in biomedical diagnostics. Additionally, the VGG 16 model is also a lightweight model with less parameters to train in comparison to other models. The two major findings of this study are:
\begin{itemize}
    \item For highly noised breast ultrasound images, clustering-based unsupervised segmentation with K-Means++ degrades the performance of semantic segmentation model U-Net, whereas super-pixel-based unsupervised segmentation with SLIC complements the performance of semantic segmentation model U-Net.
    \item In breast cancer diagnosis from breast ultrasound images, despite being a lightweight integrated architecture with fewer parameters to train, SLIC-UNET-VGG16 is outperforming as a deep integrated pipeline based on statistical evaluation.
\end{itemize}

However, as stated in the Section \ref{literature-review}, in previous state of the art studies, researchers separately worked with segmentation or classification of breast ultrasonography, histopathology and MRI images. But to ensure a complete automated diagnosis which helps the medical practitioners to provide an early diagnosis at low cost a complete integrate pipeline of breast cancer diagnosis is necessary. An end to end integrated pipeline is crucial for industry level medical application to support breast cancer diagnosis. And so, considering the potential findings and rigorous analysis, we propose a deep integrated model including SLIC (Simple Linear Iterative Clustering) as a part of pre-processing and Modified version of U-Net for semantic segmentation leading to pretrained VGG 16 model for feature extraction for classification purpose and in the end a combination of fully connected layers for Breast Cancer Detection from Breast Ultrasonography Images.

\section{Conclusion}
For decades, breast cancer has been a source of great concern among women all over the world. Every year, many potential lives are disrupted by Breast Cancer, and many of them die prematurely. Breast cancer detection at an early stage is a significant challenge in the field of cancer research. In this study, we explored the possibilities of Deep Learning and Computer Vision methods to analyze Breast Ultrasonography images in order to detect Breast Cancer at an early stage in a fully automated manner. However, Breast Ultrasongography Images are highly noised and complex in terms of pixel intensity distribution for ROI (Region of Interest) and Non-ROI (Other areas than ROI). To tackle this issue, we proposed a deep integrated automated pipeline, including Simple Linear Iterative Clustering (SLIC) based preprocessing followed by Semantic Segmentation with a modified version of U-Net leading to classification with Pretrained ConvNet Feature Extractor based Neural Network. However, based on statistical analysis of performance, the pre-trained VGG16 model proved to be a better feature extractor than other state of the art pre-trained models as part of the proposed integrated model. In future, we would like to collect more and more representative Breast Ultrasonography Data to train our models more effectively. Deep generative models, such as GAN (Generative Adversarial Network), have a high potential for augmenting data to tackle data scarcity problems in small datasets, complementing deep learning-based classification. So, we would integrate Generative Adversarial Network based data augmentation in this deep integrated pipeline to alleviate the performance of our proposed deep integrated framework. Furthermore, transformer-based attention models, which were previously popular for Natural Language Processing challenges, are currently being applied in the Computer Vision area due to their less complex nature and faster processing capability when compared to convolutional layers. In future, we will also explore the possibility of integrating attention based transformer models including Vision Transformer in our deep integrated framework to improve the accuracy of breast cancer diagnosis. Our proposed deep integrated pipelining framework could be applied efficiently in various bio-medical industries to solve additional issues in cancer research via automated detection, supporting medical practitioners in protecting humanity from the assault of devastating malignancies.


% \bibliographystyle{unsrtnat}
% \bibliography{references}  %%% Uncomment this line and comment out the ``thebibliography'' section below to use the external .bib file (using bibtex) .





% \end{thebibliography}

\end{document}